\documentclass[showpacs,epsfig,preprintnumbers,amsmath,amssymb,nofootinbib,20pt]{revtex4-1}
\usepackage{graphicx}
\usepackage{dcolumn}
\usepackage{bm}

\begin{document}
\def\d{{\rm d}}
\def\Epos{E_{\rm pos}}
\def\ap{\approx}
\def\eff{{\rm eft}}
\def\L{{\cal L}}
\newcommand{\vev}[1]{\langle {#1}\rangle}
\newcommand{\CL}   {C.L.}
\newcommand{\dof}  {d.o.f.}
\newcommand{\eVq}  {\text{EA}^2}
\newcommand{\Sol}  {\textsc{sol}}
\newcommand{\SlKm} {\textsc{sol+kam}}
\newcommand{\Atm}  {\textsc{atm}}
\newcommand{\Chooz}{\textsc{chooz}}
\newcommand{\Dms}  {\Delta m^2_\Sol}
\newcommand{\Dma}  {\Delta m^2_\Atm}
\newcommand{\Dcq}  {\Delta\chi^2}
\newcommand{\nbb}{$\beta\beta_{0\nu}$ }
\newcommand {\be}{\begin{equation}}
\newcommand {\ee}{\end{equation}}
\newcommand {\ba}{\begin{eqnarray}}
\newcommand {\ea}{\end{eqnarray}}
\def\VEV#1{\left\langle #1\right\rangle}
\let\vev\VEV
\def\e6{E(6)}
\def\10{SO(10)}
\def\21{SA(2) $\otimes$ U(1) }
\def\321{$\mathrm{SU(3) \otimes SU(2) \otimes U(1)}$ }
\def\lr{SA(2)$_L \otimes$ SA(2)$_R \otimes$ U(1)}
\def\422{SA(4) $\otimes$ SA(2) $\otimes$ SA(2)}
\newcommand{\AHEP}{%
School of physics, Institute for Research in Fundamental Sciences
(IPM)\\P.O.Box 19395-5531, Tehran, Iran\\

  }
\newcommand{\Tehran}{%
School of physics, Institute for Research in Fundamental Sciences (IPM)
\\
P.O.Box 19395-5531, Tehran, Iran}
\def\roughly#1{\mathrel{\raise.3ex\hbox{$#1$\kern-.75em
      \lower1ex\hbox{$\sim$}}}} \def\lsim{\roughly<}
\def\gsim{\roughly>}
\def\ltap{\raisebox{-.4ex}{\rlap{$\sim$}} \raisebox{.4ex}{$<$}}
\def\gtap{\raisebox{-.4ex}{\rlap{$\sim$}} \raisebox{.4ex}{$>$}}
\def\lsim{\raise0.3ex\hbox{$\;<$\kern-0.75em\raise-1.1ex\hbox{$\sim\;$}}}
\def\gsim{\raise0.3ex\hbox{$\;>$\kern-0.75em\raise-1.1ex\hbox{$\sim\;$}}}



\title{Constraining secret gauge interactions of neutrinos by meson decays }

\date{\today}
\author{P. Bakhti}\email{pouya\_bakhti@ipm.ir}
\author{Y. Farzan}\email{yasaman@theory.ipm.ac.ir}
\affiliation{\Tehran}
\begin{abstract}
Secret coupling of neutrinos to a new light vector boson, $Z'$, with a mass smaller than 100~MeV is motivated within a myriad of scenarios which are designed to explain various anomalies in particle physics and cosmology. Due to the longitudinal component of the massive vector boson, the rates of three-body decay of charged mesons ($M$) such as the pion and the kaon to the light lepton plus neutrino and $Z'$ ($M \to l \nu Z'$) are enhanced by a  factor of $(m_M/m_{Z'})^2$. On the other hand, the standard two body decay  $M \to l \nu$  is suppressed by a factor of $(m_l/m_M)^2$ due to chirality.  We show that in the case of ($M \to e \nu Z^\prime$), the enhancement of $m_M^4/m_e^2 m_{Z^\prime}^2\sim 10^8-10^{10}$ relative to two-body decay  ($M \to e \nu$) enables us to probe very small values of gauge coupling for $\nu_e$.   The strongest bound comes from the $R_K\equiv Br(K \to e +\nu)/Br(K \to \mu +\nu)$ measurement in the NA62 experiment.
The bound can be significantly improved by customized searches for signals of three-body charged meson decay into the positron plus missing energy in the NA62 and/or PIENU data.
\end{abstract}
{}
\date{\today}
\maketitle
\section{Introduction}
As it is well-known, although the energy frontier CMS and ATLAS collaborations at the LHC have discovered the Higgs and measured its mass and most important couplings
with a remarkable precision, there is no sign of  much-sought-after new physics beyond the Standard Model (SM). On the other hand, the neutrino physics program and various kaon decay experiments in the
luminosity frontier are making fast progress. It is intriguing to speculate about the possibility of interaction of neutrinos with new light particles, especially introducing a new  gauge interaction which involves neutrinos without  a coupling to charged leptons ( i.e. in contrast to electromagnetic interactions that involve charged leptons but not the neutrinos.) Reference  \cite{Farzan:2016wym}
proposes a scenario for this pattern of interaction by introducing a new  fermion of mass $\sim 1$~GeV charged under new $U(1)^\prime$ gauge symmetry and mixed with $\nu_\alpha$ with a mixing $\kappa_\alpha$. Let us denote the new gauge boson with $Z'$ and its gauge coupling with $g_{NEW}$.  As shown in  \cite{Farzan:2016wym}, the active neutrinos will obtain interactions of form
$g_{NEW} \kappa_{\alpha }\kappa^*_{\beta } \bar{\nu}_\beta \gamma^\mu \nu_\alpha Z'_\mu$.

Such neutrinophilic new gauge interaction with a light gauge boson is motivated by the so-called  $\nu$DM models which are proposed to solve small scale structure problems that appear in canonic collisionless  cold dark matter paradigm \cite{Chu:2015ipa,Dasgupta:2013zpn,Zurek,Shoemaker:2013tda,Aarssen:2012fx,Bringmann:2006mu,Boehm:2000gq}. Within these models, both  dark matter particles and neutrinos (but not other SM particles) enjoy a new gauge interaction. As a result, the DM particles can interact with each other solving the cusped-cored problem and moreover neutrinos and dark matter particles in early universe can scatter off each other via the new gauge interaction, leading to late time kinetic dark matter decoupling and therefore increasing the cut-off in the structure power spectrum.

Decays of charged mesons can reveal secrets of neutrinos interacting with light new particles. Kaon decays have been widely used to constrain the Majoron  \cite{Farzan:2016wym,Britton:1993cj,Picciotto:1987pp,Gelmini:1982rr,Barger:1981vd,Pasquini:2015fjv,Lessa:2007up} and SLIM model \cite{Farzan:2010wh}. Moreover ($K\to \mu +{\rm missing ~energy}$)  have been used to constrain the gauge coupling of muon to a light gauge boson which is motivated as a solution to the $(g-2)_\mu$ anomaly. In this article, we focus on the scenario that only neutrino couples to the new gauge boson as motivated by \cite{Chu:2015ipa,Dasgupta:2013zpn,Hannestad:2013ana,Ho:2012br,Tang:2014yla}. Thus, at the tree level, $Z'$ can only decay into neutrinos and hence appears as missing energy in the experiments.  While standard two-body decay $M \to \nu l$ is suppressed by $m_l^2/m_M^2$, for light $Z'$ the three-body decay $M \to l \nu  Z'$ from longitudinal polarization of $Z'$ receives an enhancement of $m_M^2/m_{Z'}^2$. Thus, if the new gauge coupling is not too small, we expect a significant contribution to ($\pi^+ {\rm or}~ K^+\to e^+ +{\rm missing~ energy}$). This huge enhancement of $O\left(m_K^4/m_e^2 m_{Z^\prime}^2\right)\sim 10^8-10^{10}$ brings about an opportunity of probing very small gauge couplings of $\nu_e$ which has been mostly overlooked, being overshadowed by a wide interest in the gauge interaction of second-generation leptons for which the enhancement is much weaker.

The aim of the present paper is to investigate the sensitivity of leptonic decay of charged mesons to interaction of neutrinos with new light gauge boson. We study  the implications of the three-body ($\pi^+\to e^++{\rm missing~ energy}$) data from old TRIUMF data \cite{Picciotto:1987pp}, measurement of ($K^+ \to e^++\nu\nu\nu$) \cite{Heintze:1977kk}, measurement of $R_\pi\equiv \Gamma (\pi^+ \to e^+ +\nu)/\Gamma (\pi^+ \to \mu^+ +\nu)$ by the PIENU experiment \cite{Aguilar-Arevalo:2015cdf} and measurement of  $R_K\equiv \Gamma (K^+ \to e^+ +\nu)/\Gamma (K^+ \to \mu^+ +\nu)$ by KLOE \cite{AmelinoCamelia:2010me} and NA62 \cite{Lazzeroni:2012cx}.
 We find that  the NA62  measurement of $R_K$ \cite{Lazzeroni:2012cx} yields the strongest bound on the new gauge interaction of $\nu_e$. For completeness, we also study similar bounds on new gauge coupling of $\nu_\mu$ from  $Br (K^+ \to \mu^+ \nu \nu \nu)<2.4 \times 10^{-6}$ at 90 \% C.L. \cite{Artamonov:2016wby}.

In Sec. II, we present formulas for the decay rate for $M\to l \nu Z^\prime$. In Sec. III, bounds from various meson decays are discussed. The results are summarized in Sec. IV.

\section{Meson decay}
\begin{figure}
\begin{center}
        \includegraphics[width=0.3\textwidth]{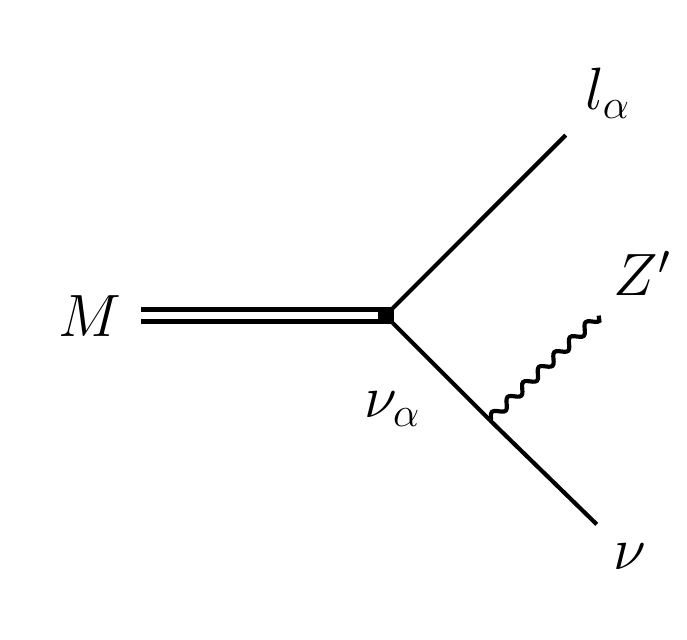}
 \end{center}

\caption{\label{feyn} Feynman diagram for charged meson decay to a charged lepton, $l_\alpha$, neutrino and the new light vector boson, $Z'$.}
\end{figure}

 Let us write the interaction of the neutrino of active flavor $\alpha$ with the new vector boson as follows
 \be g_{\alpha i} Z'_\mu \bar{\nu}_i \gamma^\mu \nu_\alpha \label{gauge-int}\ee
 where $\nu_i$ are any neutrino mass eigenstates  much lighter than 100 MeV ( i.e. pion mass). Within the scenario proposed in Ref. \cite{Farzan:2016wym} and described in the Introduction, $g_{\alpha i}=\sum_\beta \kappa_\alpha \kappa^*_\beta U_{\beta i}^*$ but $\nu_i$ can, in general, also involve new light mass eigenstates  which should be  dominantly composed of  sterile neutrinos.
 As shown in Fig \ref{feyn}, this coupling leads to charged meson decay with
\begin{equation}\label{decayrate}
\Gamma(M\longrightarrow l_\alpha\nu Z')=\frac{1}{64\pi^3m_M}\int_{E_l^{min}}^{E_l^{max}}\int_{E_\nu^{min}}^{E_\nu^{max}}  dE_l  dE_\nu \sum_{spins}\vert {\cal M} \vert^2.
\end{equation}
Neglecting the neutrino and lepton masses, we can write \be \label{Msquared}\sum_{spins}\vert {\cal M} \vert^2=  (\sum_i g_{\alpha i}^2)G_F^2f_M^2V_{qq'}^2\left(m_M^2+m_{Z'}^2-2m_ME_{Z'}+\frac{(m_M^2-m_{Z'}^2-2m_ME_l)(m_M^2-m_{Z'}^2-2m_ME_\nu)}{m_{Z'}^2}\right),
\ee
in which $V_{qq'}$ and  $f_M$ are the relevant CKM mixing element and meson decay constant, respectively.
The integration limits, again neglecting the neutrino and lepton masses,  are
$$E_l^{min}=m_l,~~~~~~~~~~~~~~~~~~~~~~~~~~~~~~~~~~~~ E_l^{max}=\frac{m_M^2-m_{Z'}^2}{2m_M},$$
$$E_\nu^{min}=\frac{m_M^2-m_{Z'}^2-2m_ME_l}{2m_M},~~~~~~~~~ E_\nu^{max}=\frac{m_M^2-m_{Z'}^2-2m_KE_l}{2(m_M-2E_l)}.$$
Neglecting neutrino masses is, of course,  justified. Neglecting the mass of the final charged lepton causes a correction of $O(m_l^2/m_M^2)$.  In case of decay into electron and positron, the correction is less than $O(10^{-5})$ and completely negligible. Calculating the rate of kaon decay into the muon and missing energy, we have kept the muon mass which induces a correction of 5$\%$.

 As mentioned above, $\nu_i$ in Eq. (\ref{gauge-int}) can also denote a new light neutrino state such as the hypothetical sterile neutrinos with mass $O(1~{\rm eV})$ that provide the famous 3+1 solution to the LSND anomaly or the reactor and Gallium anomalies.
  A light sterile neutrino with mass $O(1~{\rm eV})$ and mixing of $O(0.1)$ provides the most popular solution for these anomalies but is disfavored by cosmological bounds. However, it has been recently shown that if the sterile neutrino has a gauge interaction with light gauge boson these bounds can be relaxed
\cite{Chu:2015ipa,Dasgupta:2013zpn,Hannestad:2013ana,Ho:2012br,Tang:2014yla}. Reference \cite{Chu:2015ipa} combines various bounds from the big bang nucleosynthesis, cosmic microwave background and large scale structure and finds two distinct regions  in the parameter space $(m_{Z'}, g_{NEW})$ which are compatible with all the bounds and accommodate a sterile neutrino suitable for solving the LSND anomaly. Notice, however, that in order to obtain interaction between $\nu_\alpha$ and $Z'$, it is required that the unitarity of the mixing (sub)matrix of light states is violated. If no heavy state with mass higher than $\sim 100$ MeV is introduced, the unitarity of the mixing matrix will forbid the active neutrino coupling to $Z'$. We may consider the scenario  where both the active ($\nu_a$) and light sterile neutrino ($\nu_s$ with mass eV) are neutral under new $U(1)^\prime$ and, therefore, mixing between $\nu_a$ and $\nu_s$ can be  obtained with ordinary Yukawa coupling without a need to  break $U(1)^\prime$ and $\nu_a$ and $\nu_s$  both mix with the heavy state charged under the new $U(1)^\prime$ with mixings $\kappa_a$ and $\kappa_s$. We shall, therefore, obtain interactions of form $ g_{NEW} \kappa_\alpha^*\kappa_\beta \bar{\nu}_\alpha \gamma^\mu \nu_\beta Z'_\mu$, $ g_{NEW} \kappa_s^*\kappa_\beta \bar{\nu}_s \gamma^\mu \nu_\beta Z'_\mu$ and $ g_{NEW} |\kappa_s|^2 \bar{\nu}_s \gamma^\mu \nu_s Z'_\mu$,
where the first two can lead to a signal at meson decay and the third can relax the tension with cosmology as described in Refs. \cite{Chu:2015ipa,Dasgupta:2013zpn,Hannestad:2013ana,Ho:2012br,Tang:2014yla}.

Up to now we have assumed that only SM particles that (at tree level) interact with $Z'$ are neutrinos so $Z'$ can only decay into neutrinos appearing as missing energy.
If, as postulated in Ref. \cite{Farzan:2016wym}, $Z'$ also couples to quarks as long as $m_{Z'}<100$ MeV, still the only decay mode kinematically available for $Z'$ is decay into neutrinos and a discussion similar to the above applies. One should, however, also include the contribution from the quark-$Z'$ interaction to the three-body meson decay. For the case like \cite{Farzan:2016wym} where the coupling to the quark is given by the baryon number, there is no mesonic internal bremsstrahlung contribution but an effect from virtual baryons is expected.

\section{Bounds}

The bounds on the coupling of $\nu_e$ to the new gauge coupling are summarized in Fig. 2. Decreasing $m_{Z^\prime}$ the bound becomes stronger.
This is understandable because $Br(M \to Z^\prime l\nu)$ increases with decreasing $m_{Z^\prime}$ [see Eq. (\ref{Msquared})]. As shown in the figure, the strongest bound comes from
the $R_K$ measurement at NA62.  $R_K$ is defined as
\be R_K\equiv \frac{{\rm Br}(K \to e \nu)}{{\rm Br}(K \to \mu \nu)}\label{RK}\ee
and is traditionally considered an excellent measure to test lepton flavor universality because its prediction is free from uncertainties in the Kaon form factor. The standard model prediction, taking into account the internal bremsstrahlung emission, is \cite{Cirigliano:2011ny,Cirigliano:2007xi}
$$R^{SM}_K=(2.477 \pm 0.001)\times 10^{-5}.$$
To extract $Br( K \to e \nu)$, the signal for $(K \to e+{\rm missing~ energy})$ with $(P_K-P_e)^2 \simeq 0$ is collected. In recent years in the energy frontier, experiments such as KLOE II \cite{AmelinoCamelia:2010me}, NA48 \cite{Batley:2007aa},  NA62 \cite{Lazzeroni:2012cx} and E494 \cite{Artamonov:2009sz} have studied kaon decay with an unprecedented accuracy.
 The NA62 experiment \cite{Lazzeroni:2012cx,pdg} which provides the best measurement finds that for $0.013~{\rm GeV}^2<(P_K-P_e)^2$, $$R_K^{EXP}=(2.488 \pm 0.010)\times 10^{-5}.$$

The Particle Data Group average is also very close to Eq. (6) \cite{pdg} . Notice that there is a small 1~$\sigma$ excess relative to the SM prediction. In our model, there will be a new positive contribution to $Br( K\to e+{\rm missing~ energy})|_{0.013~{\rm GeV}^2<(P_K-P_e)^2}$ (and, therefore, to $R_K$) which can be obtained from Eq. (\ref{decayrate}) when setting the lower bound of the integration of $E_l$ to 233.68 MeV (in order to obtain $(P_K-P_e)^2>0.013~{\rm GeV}^2$). We have defined $$\chi_K^2 =\frac{(R_K^{EXP}-R_K^{SM}-R_K^{NEW})^2}{\sigma^2}$$ where $R_K^{NEW}$ is the contribution from $K\to e Z^\prime \nu$ with $E_e> 233.68$ MeV at the kaon rest frame and $\sigma=0.010\times 10^{-5}$ is the NA62 uncertainty.  The theoretical uncertainty in the SM prediction of $R_K$, being 1 order of magnitude smaller is neglected.
 Setting $\chi_K^2<2.71$, we have found the 90 \% C.L. bound shown in Fig. 2. Notice that the best fit corresponds to nonzero value $ \sqrt{\sum_i g_{i e}^2}=0.004(m_{Z^\prime}/30~{\rm MeV})$.
 Similar analysis of the latest $R_K$ measurement at KLOE \cite{Ambrosino:2009aa} leads to a bound which is weaker by a factor of 4.

\begin{figure}[t] \centering
	\includegraphics[width=0.6\textwidth]{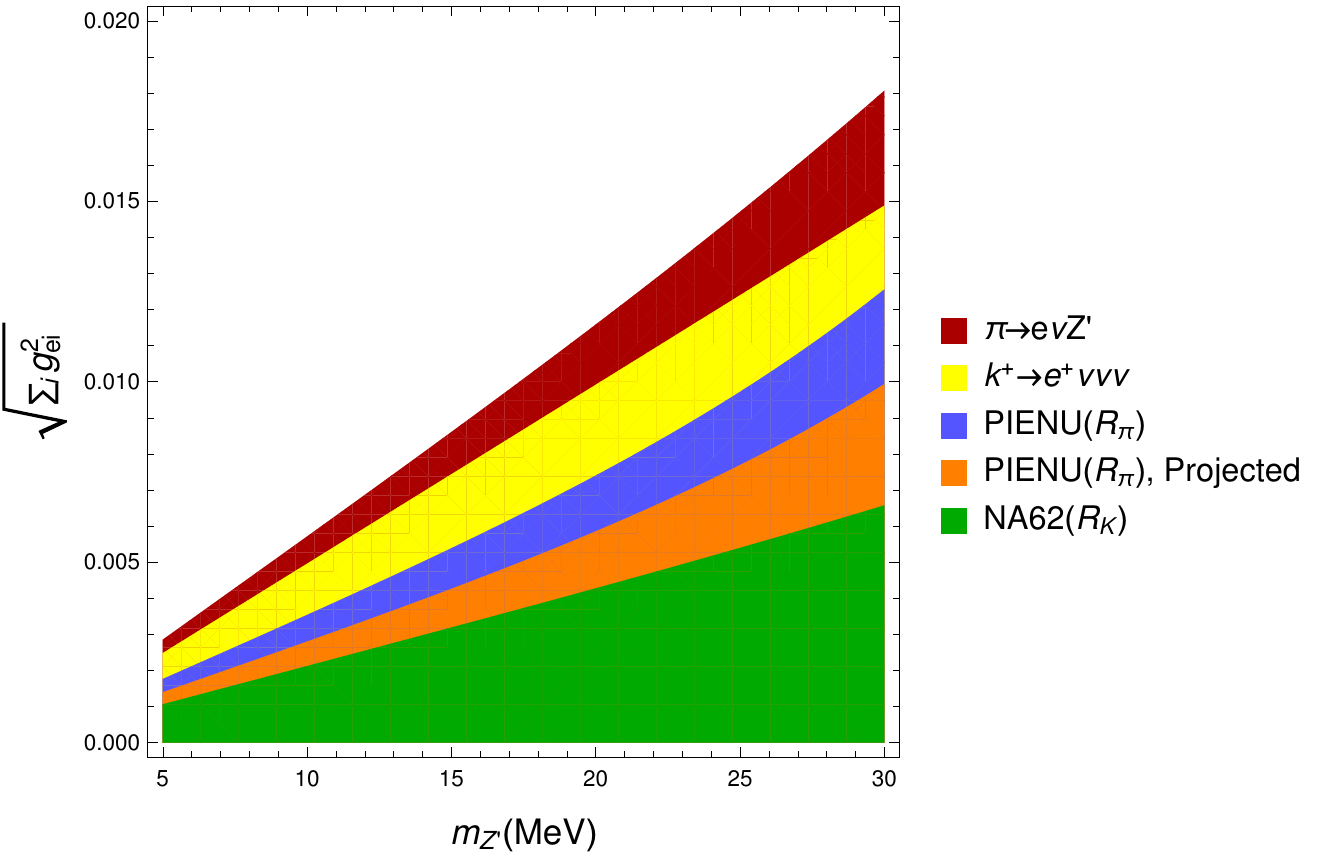}
       \caption{\label{Ei} The 90$\%$ C.L. constraints on   $\sqrt{\sum_i g_{ei}^2}$ versus $m_{Z^\prime}$ from constraints on $\pi\longrightarrow e\nu Z'$ \cite{Picciotto:1987pp}  and  $K^+\longrightarrow e^+\nu\nu\nu$ \cite{Heintze:1977kk}  branching ratios, from current and projected $R_\pi$  measurement by PIENU \cite{Aguilar-Arevalo:2015cdf},  and from the $R_K$ measurement by  NA62 \cite{Lazzeroni:2012cx}.}
\end{figure}

\begin{figure}[t] \centering
	\includegraphics[width=0.5\textwidth]{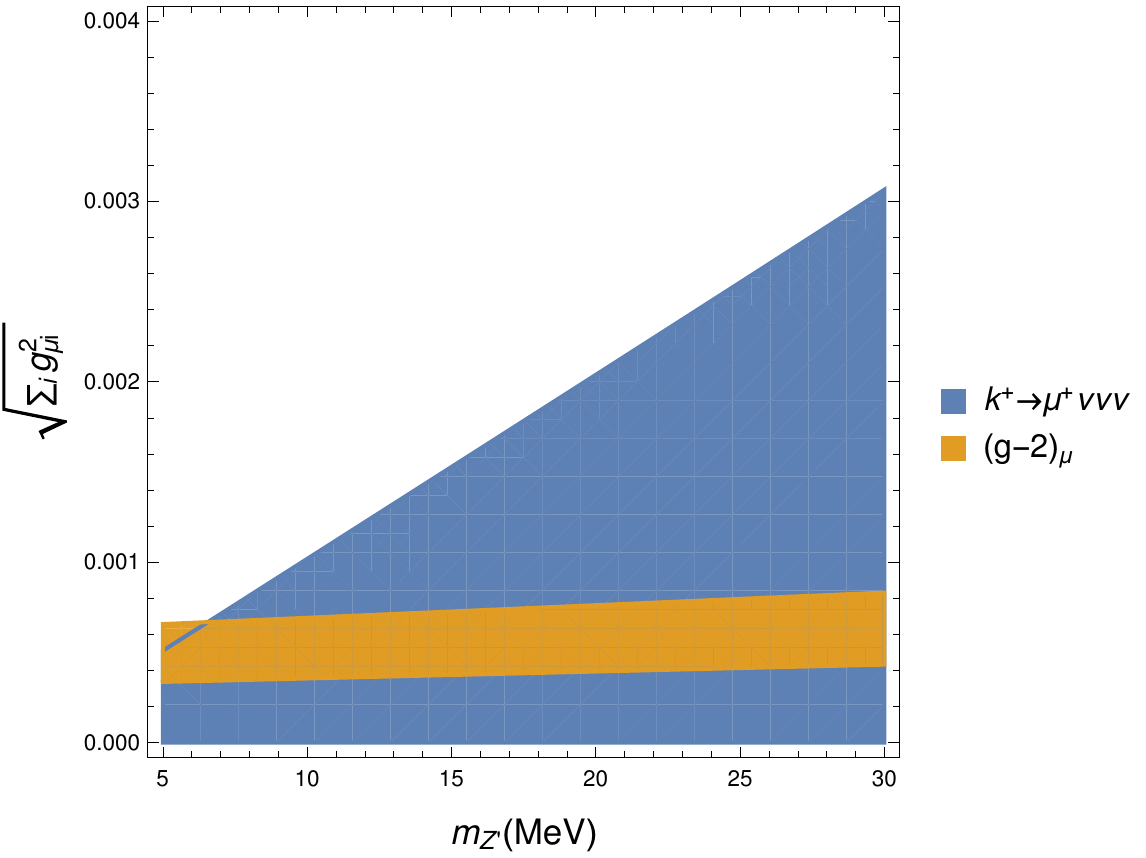}
        \caption{\label{Ei1} The 90$\%$ C.L. constraints on the  $\sqrt{\sum_i g_{\mu i}^2}$  versus $m_{Z^\prime}$ from $K^+\longrightarrow \mu^+\nu\nu\nu$  branching ratio \cite{Artamonov:2016wby}. The band shows the parameter space within the   $L_\mu$ gauge models (giving rise to equal couplings to $\mu$ and $\nu_\mu$) that can explain the $(g-2)_\mu$ anomaly \cite{Pospelov:2008zw}. }
\end{figure}

Replacing $K$ meson with $\pi$ meson in Eq. (\ref{RK}) we arrive at the definition of  $R_\pi$. Similarly to $R_K$, $R_\pi$ does not depend on form factors and its prediction in the standard model does not suffer from uncertainties in the value of form factor so it is a suitable tool to test new physics. The SM prediction, including soft photon radiation, is \cite{Marciano:1993sh,Cirigliano:2007ga,Bryman:2011zz}
$$R^{SM}_\pi =(1.2352\pm 0.0002)\times 10^{-4}.$$ The TRIUMF PIENU experiment \cite{Aguilar-Arevalo:2015cdf} has measured $R_\pi$ by applying lower cut of 52~MeV on the electron energy. Analyzing the data taken by 2010
\be R_\pi^{EXP} =[1.2344\pm 0.0023 (stat)\pm 0.0019 (sys)] \times 10^{-4}.\label{pie}\ee It is projected that by analyzing the rest of the data taken before 2012, the statistical uncertainty can be reduced by a factor of 3 \cite{ITO}. Within our model, there will again be a positive contribution to $R_\pi$ from $\pi \to e Z^\prime \nu$ with $E_l>52$ MeV. Similarly to $\chi_K^2$, we have defined $\chi^2_\pi$ with
$\sigma^2=\sigma_{stat}^2+\sigma_{sys}^2$. Again the theoretical uncertainty in the $R_\pi$ prediction is negligible. The present (with $\sigma_{stat}=2.3\times 10^{-7}$) and projected (with $\sigma_{stat}=0.7\times 10^{-7}$) bounds from PIENU are found by setting $\chi^2_\pi<2.71$ and are displayed in Fig 2. They are slightly weaker than the bound from NA62. Notice that  even within the SM, $M \to l \nu \gamma$ with soft photon which escapes detection can contribute to $R_M$ (i.e., $R_\pi$ or $R_K$) . The SM predictions quoted above take into account this effect for each setup.

Combining the results of various experiments, PDG \cite{pdg} reports $R_\pi^{PDG}=(1.2327\pm 0.0023)\times 10^{-4}$. If instead of Eq. (\ref{pie}), we used the PDG average, we would find a bound as strong as the one from NA62; however, one should take this bound with a grain of salt for two reasons: (i) The main reason why the bound is so strong is that the PDG average is more than one sigma below the SM prediction which is most likely due to a statistical fluctuation and will change with further statistics, and (ii) Various experiments may have used different cuts on positron energy to extract $Br(\pi^+ \to e^+ \nu)$. Such difference can have nontrivial consequences for our analysis.

 From \cite{Picciotto:1987pp}, we know that at 90 \% C.L., 
 \be R=\frac{\Gamma( \pi \to e \nu Z')}{\Gamma (\pi \to \mu \nu)}< 4 \times 10^{-6} \ee
 which can be translated into a bound on $(\sum_i |g_{ei}|^2)^{1/2}$ as shown in Fig 2.  Notice that Refs.  \cite{Picciotto:1987pp,pdg} were, in fact, interested in the Majoron emission with mass in the range (0-125) MeV but a signal for $\pi \to e \nu Z'$ will be exactly similar. Figure 2 also shows the limit from the bound $Br (K^+ \to e^+ \nu \nu \nu)<6 \times 10^{-5}$ at 90 \% C.L. \cite{pdg} interpreted as $Br(K^+ \to e^+ \nu Z')$. As seen in the figure, although these bounds  from old data taken in the 1980s are less restrictive  than the bounds from the recent measurement of $R_K$ and $R_\pi$ but they are still comparable. That is because in the case of $R_K$ and $R_\pi$, cuts have been applied to select two-body decay. If the full data are analyzed with customized searches for three-body decay $M \to l\nu Z^\prime$, a larger part of parameter space can, of course, be probed.

 For  completeness, we also show the bound from $K \to \mu \nu \nu \nu$ on the new coupling of $\nu_\mu$ in Fig. 3. Similar analysis has been carried out in the literature for the model where the new vector
 boson couples to the right-handed muon \cite{Barger:2011mt}, the left-handed neutrino \cite{Belotsky:2001fb} or the left-handed $\mu$ and $\nu$ \cite{Laha:2013xua,Ibe:2016dir}.  For comparison, we show $2 \sigma$ band explaining $(g-2)_\mu$ for the case that left-handed $\mu$ couples to the new vector boson \cite{Pospelov:2008zw}. The bound from $(g-2)_\mu$ does not apply to our model in which only neutrinos have the new interaction at tree level.

 We already mentioned that if the $g_{\alpha \beta} \bar{\nu}_\alpha \gamma^\mu \nu_\beta Z^\prime_\mu$ coupling comes from the mixing ($\kappa_\alpha$) of $\nu_\alpha$ with a heavier fermion charged under new $U(1)^\prime$, the coupling is given by $\kappa_\alpha \kappa_\beta g_{NEW}$. From the violation of unitarity \cite{Fernandez-Martinez:2016lgt}, we know $|\kappa_e|^2<2.5 \times 10^{-3} $ and $|\kappa_\mu|^2<4.4 \times 10^{-4} $ so within this scenario, the NA62 bound restricts $g_{NEW}<0.4 (m_{Z^\prime}/5~{\rm MeV})$.
 Notice that the big bang nucleosynthesis sets a lower bound of $\sim 5$ MeV on the mass of $Z^\prime$ which is in thermodynamic equilibrium with neutrinos \cite{Kamada:2015era}. For this reason, we have cut our figures at 5 MeV.

 \section{Conclusions and outlook}
We have studied the bounds from  pion and kaon decay on coupling of neutrinos to a light new vector boson. We have found that the strongest bound on the coupling of $\nu_e$ comes from the $R_K$ measurement at NA62 experiments which at 90 \% C.L. is $(\sum_i g_{ei}^2)^{1/2}<0.0067~(m_{Z^\prime}/30~{\rm MeV})$. A slightly weaker bound comes from the $R_\pi$ measurement by PIENU experiment.
To extract $R_M$, both experiments have applied cuts on the energy of $e^+$ to reduce the background to two-body decay signal $M^+\to e^+ \nu$. Since the decay mode of interest is three-body, we expect a customized analysis by the collaboration searching for $[K~({\rm or}~\pi) \to e \nu Z^\prime ]$ can probe a much wider range of parameters (down to $g^\prime \sim 0.001 (m_{Z^\prime}/30~{\rm MeV})$).
 \section*  {Acknowledgements}
 We  are grateful to G Ruggiero for useful comments and encouragement and also thank A. Abada, F. Mahmoudi, A Arbey and M Bahrami-Nasr for useful discussions in the early stages of this work.
 We especially thank A. Shaikhiev for useful comments and information on experimental updates.
This project has received funding from the European Union's Horizon 2020 research and innovation programme under the Marie Sk\l{}odowska-Curie Grant Agreement No.~674896 and No.~690575.
Y. F. is also grateful to the ICTP associate office and Iran National Science Foundation (INSF) for partial financial support under Contract No. 94/saad/43287. We also thank   LPT of Orsay Univesity, where a part of this work was completed, for their hospitality. P. B. is grateful to T. Toma for useful discussions in the early stages of this work.


\end{document}